# Universal Relations for Solitary Waves in Granular Crystals under Finite Rise-decay Duration Shocks

M. Arif Hasan, Sia Nemat-Nasser

Center of Excellence for Advanced Materials

Department of Mechanical and Aerospace Engineering

Jacobs School of Engineering

University of California, San Diego

**ABSTRACT**

We focus on solitary waves generated in arrays of lightly contacting spherical elastic granules by shock forces of steep rise and slow decay durations, and establish *a priori*: (i) whether the peak value of the resulting solitary wave would be greater, equal, or less than the peak value of the input shock force; (ii) the magnitude of the peak value of the solitary waves; (iii) the magnitude of the linear momentum in each solitary wave; (iv) the magnitude of the linear momentum added to the remaining granules, if the first granule is ejected; and (v) a quantitative estimate of the effect of the granules' radius, density and stiffness on force amplification/mitigation. We have supported the analytical results by direct numerical simulations.

## 1. Introduction

Ordered granular media are strongly nonlinear phononic crystals [1] highly suited for stress-wave management [2]. They may have one-, two-, or three-dimensional alignments [3–10], where the nearest neighbors exchange energy and momentum through strongly nonlinear Hertzian interaction [11]. Nesterenko is the first to show that such a nonlinear relation between contact force and position produces a new class of *translational solitary waves that involve no vibrations* and have a compact support of about 5 granules [3,4] in monodispersed granular chains. They are excellent media for passive energy absorption and shock mitigation [3]. Diatomic and tapered chains are also studied by several researchers analytically, numerically, and experimentally [4,8,12–17] as shock protectors. These studies, however, consider impulsive loads. In reality, the applied forces due to impact- or explosive-induced shocks have finite steep rise and slow decay periods that have not been theoretically studied. It may also be possible to produce input forces with slow rise and steep decay periods. While our focus here is on mechanically induced shocks, our approach may also be applicable to input forces with slow rise and fast decay periods, for which one or several granules may be ejected, adding linear momentum to the remaining granules. However, we do not address this issue in the present paper.

## 2. Statement of Problem and System Description

When the applied force is an *impulse,* imparting an initial velocity to the first granule of an ordered granular crystal, then after an initial transient period, the contact force between the granules forms a single localized wave-form, called solitary wave (SW). We seek to broaden the class of input forces, focusing on realistic shock loads which generally have steep rise and slow decay finite durations.

Consider a set of lightly contacting one-dimensional identical elastic spheres, subjected to an applied force, $F(t;t_0)$, of finite rise time, $t_0$, and finite decay time.

The dimensionless governing equations of motion are

$$x_1'' = \tilde{F}(\tau) - \left[(x_1 - x_2)_+^{3/2}\right], \quad x_i'' = \left[(x_{i-1} - x_i)_+^{3/2} - (x_i - x_{i+1})_+^{3/2}\right], \quad x_N'' = \left[(x_{N-1} - x_N)_+^{3/2}\right], \quad (1)$$

where prime denotes differentiation with respect to normalized time $\tau = t/\beta$, with $t$ being the dimensional time unit, and $\beta = \left[2\sqrt{2}(1-\nu^2)\pi\right]^{-1/2}\left[\sqrt{(E/\rho)}/R\right]$, where $E$ is the Young's modulus, $\nu$ is the Poisson's ratio, and $R$ is the radius of a typical granule. In (1), $x_i$ is normalized displacement of the $i^{th}$ granule, measured relative to its initial position, $i = 2, \cdots, N-1$, $\tilde{F}(\tau)$ is the normalized externally applied force to the first granule, $\tilde{F}(\tau) = F(t;t_0)/\bar{\alpha}$, where $\bar{\alpha} = \sqrt{2}ER^2\left[3(1-\nu^2)\right]^{-1}$. In the absence of loads, the granules are in light contact with no compression, forming a system of No-tension elastic medium [18].

The considered input (dimensional) shock, $F(t;t_0)$, is defined by,

$$F(t;t_0) = F_0\left[F_1(t)H(t_0-t) + F_2(t)H(t-t_0)H(nt_0-t)\right]; \quad t \in (0,\infty], \tag{2}$$

*which has finite rise and finite decay durations,* a common characteristic of an actual shock. Here, $t_0$ is the rise time (used as unit of time), $nt_0$ is the total duration of the shock with $\{n \in \mathbb{R} \mid n > 1\}$, $F_0$ is the peak value of the shock (used as unit of force), $F_1(t)$ and $F_2(t)$ with peak values of 1 represent shock profiles during rise and decay periods, respectively, and $H(\bullet)$ is the Heaviside step function. Therefore, $F(t;t_0) = F_0 F_1(t)$ for $0 \leq t \leq t_0$, $F(t;t_0) = F_0 F_2(t)$ for $t_0 \leq t \leq nt_0$, $F(t;t_0) = 0$ for $t \geq nt_0$, and $\max|F_1(t)| = \max|F_2(t)| = 1$.

In general, impact- or explosive-induced shocks have steep rise followed by slow decay periods [19]. We focus on a class of such shocks and limit our attention to realistic cases where the input rise and decay profiles of the applied force, $F_1(t)$ and $F_2(t)$, satisfy,

$$\int_{t_0}^{nt_0} F_2(t)dt \geq \int_0^{t_0} F_1(t)dt; \quad \int_0^{t_0} F_1(t)dt \geq \frac{1}{2}t_0; \quad \int_{t_0}^{nt_0} F_2(t)dt \leq \frac{(n-1)}{2}t_0. \tag{3}$$

The first condition guarantees that the rise is faster than the decay, and the second one requires that the area under the rise profile be at least equal to that of a triangle. Finally, the last restriction ensures that the area under the decay profile is not greater than that of a triangle with the same duration.

## 3. Characteristics of SW

The linear momentum of a SW traveling in ordered granular crystals, is $L_{SW} = \int_0^\infty F_{SW}(t)dt$, where $F_{SW}(t)$ represents the time variation of SW. The profile of a SW can be expressed in terms of the Padé approximation [20] as follows:

$$F_{SW}(t) \simeq \frac{\sqrt{2RE}}{3(1-v^2)} [BS(\hat{\tau})]^{3/2}; \quad \hat{\tau} = (B/R)^{1/4} t/\beta, \tag{4a}$$

where $B$ is the peak amplitude of the relative displacement of two contacting granules, and $S(\hat{\tau})$ is given by $S(\hat{\tau}) = (1 + q_2\hat{\tau}^2 + q_4\hat{\tau}^4 + q_6\hat{\tau}^6 + q_8\hat{\tau}^8)^{-2}$, with $q_2 = 0.439$, $q_4 = 0.0994$, $q_6 = 0.0147$ and $q_8 = 0.0013$; see [20]. According to the Hertzian contact, the peak value, $F_{SW_{max}}$, of the SW is given by $\frac{\sqrt{2RE}}{3(1-v^2)} B^{3/2}$. Hence,

$$L_{SW} \simeq F_{SW_{max}} \beta \left(\frac{R}{B}\right)^{1/4} \int_{-\infty}^{\infty} \{S(\hat{\tau})\}^{3/2} d\hat{\tau} = 1.543 F_{SW_{max}} \beta \left(\frac{R}{B}\right)^{1/4} \simeq 4.06 \left(\frac{E}{(1-v^2)\rho^{3/2} R^4}\right)^{-1/3} F_{SW_{max}}^{5/6}.$$

Our numerical simulations show that the Padé approximation underestimates the value of the integral in the above expression by about 7%; see Supplemental Material. Hence the coefficient 4.06 needs to be increased by about 7% to 13/3, leading to the following fundamental result:

$$L_{SW} \simeq A F_{SW_{max}}^{5/6}, \quad A = \frac{13}{3} \left(\frac{(1-v^2)\rho^{3/2} R^4}{E}\right)^{1/3}. \tag{4b,c}$$

The parameter $A$ characterizes the total linear momentum of a fully developed SW that the corresponding chain of contacting granules can support. For a given shock input, the peak value of the resulting SW, $F_{SW_{max}}$, also depends on the parameter $A$, and, as shown later on, this peak value increases with granules' increasing stiffness and granules' decreasing density and radius, according to the structure of parameter $A$. Therefore *A embodies significant amount of information about the resulting* SW.

## 4. Response of the System

The total linear momentum of the input force (2) is $L_{Applied} \equiv \left[ \int_0^{t_0} F_1(t)dt + \int_{t_0}^{nt_0} F_2(t)dt \right] F_0$. In certain cases, the applied force may introduce a train of SWs. Then, only a part of the input momentum will be used to create the primary SW. Let the portion $L_{Applied}^P = A F_{SW_{max}}^{5/6}$ of the total input momentum, $L_{Applied}$, be used to form the primary SW of the peak amplitude $F_{SW_{max}}$, where the superscript "$P$" stands for the primary SW. Set $t^P = \alpha A F_{SW_{max}}^{-1/6}$, where $\alpha$ accounts for the profile of the input force (see comments below) and note,

$$L_{Applied}^P = \begin{cases} \left[ \int_0^{t_0} F_1(t)dt + \int_{t_0}^{nt_0} F_2(t)dt \right] F_0 = L_{Applied}; \\ \left[ \int_0^{t_0} F_1(t)dt + \int_{t_0}^{\left(\alpha A F_{SW_{max}}^{-1/6}\right)} F_2(t)dt \right] F_0 < L_{Applied}; \quad \text{if } \left(\alpha A F_{SW_{max}}^{-1/6}\right) < nt_0 \end{cases} \quad (5a)$$

The top expression corresponds to cases when the entire linear momentum is used to form a single (primary) SW, and the second expression is for cases when more than one SW would be formed. The time-parameter $t^P = \alpha A F_{SW_{max}}^{-1/6}$ can be calculated from $\int_{t^P}^{nt_0} F_2(t)dt = \left(L_{Applied} - L_{SW}\right)/F_0$. For a triangular input shock of peak value $F_0$, it can easily be shown that the time parameter $t^P$ is given by, $t^P = nt_0 - \sqrt{(n-1)t_0 \left[ nt_0 - 2AF_{SW_{max}}^{5/6} / F_0 \right]}$; see Supplemental Material for details. In this case, we must have

$$nt_0 \geq 2 \frac{A F_{SW_{max}}^{5/6}}{F_0} = 2 \frac{F_{SW_{max}}}{F_0} A F_{SW_{max}}^{-1/6} = \alpha A F_{SW_{max}}^{-1/6}, \quad (5b)$$

which provides an expression for $\alpha$, i.e. $\alpha = 2 \frac{F_{SW_{max}}}{F_0}$. For a general input shock profile, $\alpha$ must accounts for the deviation of the input shock profile from the corresponding triangular one. Then, for profiles which satisfy conditions (2), we have observed that the expression,

$$\alpha = 2\frac{F_{SW_{max}}}{F_0} - \left|\frac{1}{2} - \int_0^1 F_1(t_0\tau)d\tau\right|^{3/2} - \left|\frac{1}{2} - \frac{1}{(n-1)}\int_1^n F_2(t_0\tau)d\tau\right|^{3/2}, \tag{5c}$$

yields accurate results, which we have verified by extensive numerical simulations; see comments in subsections 4.1 and 4.2.

### 4.1. Qualitative Prediction of Peak Contact Force, $F_{SW_{max}}$

We now examine the qualitative relation between the peak values of the input force and that of the resulting primary SW. If the peak value of the primary SW equals $F_0$, then, according to equation (4b), the portion of the input linear momentum used to form the primary SW would equal $A F_0^{5/6}$. But, if this portion is greater (less) than $A F_0^{5/6}$, then the peak value of the primary SW would by necessity be greater (less) than $F_0$. Therefore we can make the following qualitative prediction:

i. The steady-state peak compressive contact force, $F_{SW_{max}}$, would relate to the input shock amplitude, $F_0$, in the following manner:

   a. If $\left(\alpha^{F_0} A F_0^{-1/6}\right) < nt_0$, then a train of SWs form and

$$F_{SW_{max}} \begin{cases} > F_0 \\ \simeq F_0 \\ < F_0 \end{cases} \text{if} \left(\int_0^{t_0} F_1(t)dt + \int_{t_0}^{\left(\alpha^{F_0} A F_0^{-1/6}\right)} F_2(t)dt\right) \begin{cases} > AF_0^{-1/6} \\ \simeq AF_0^{-1/6} \\ < AF_0^{-1/6} \end{cases}. \tag{5d}$$

   b. Else, a single SW forms and

$$F_{SW_{max}} \begin{cases} > F_0 \\ \simeq F_0 \\ < F_0 \end{cases} \text{if} \left(\int_0^{t_0} F_1(t)dt + \int_{t_0}^{nt_0} F_2(t)dt\right) \begin{cases} > AF_0^{-1/6} \\ \simeq AF_0^{-1/6} \\ < AF_0^{-1/6} \end{cases}, \tag{5e}$$

where $\alpha^{F_0}$ is given by,

$$\alpha^{F_0} = 2 - \left|\frac{1}{2} - \int_0^1 F_1(t_0\tau)d\tau\right|^{3/2} - \left|\frac{1}{2} - \frac{1}{(n-1)}\int_1^n F_2(t_0\tau)d\tau\right|^{3/2}.$$

Table 1 shows the qualitative comparison between $F_{SW_{max}}$ and $F_0$ for different input shock profiles, supporting the predictions based on equations (5d,e).

## 4.2. Quantitative Estimate of Peak Contact Force, $F_{SW_{max}}$

If all of the input energy and momentum are not used to form a single SW, secondary or even tertiary SWs may be generated.

### 4.2.1. Estimate of $F_{SW_{max}}$ when $L^P_{Applied} < A F_0^{5/6}$ and $t^P = nt_0$

In this case the total momentum of the input force would be used to create a single SW, and in addition the first granule would be ejected, imparting additional linear momentum to the system. (The following calculations do not include this effect, and hence the resulting estimates need to be corrected, as detailed below in subsection 4.2.2). From equation (5a) we have $L^P_{Applied} = L_{Applied} \simeq L_{SW} \simeq A F_{SW_{max}}^{5/6} < A F_0^{5/6}$ which leads to,

$$F_{SW_{max}} \simeq \left( \frac{L^P_{Applied}}{A} \right)^{6/5} < F_0. \tag{5f}$$

This is a relation between $L^P_{Applied}$ and $F_{SW_{max}}$. It provides a *quantitative* estimate of the influence of the properties of the granules on the peak-force amplification/mitigation, for finite duration shock loads, showing how the peak value of the SW increases with increasing stiffness and decreasing density and radius of the granules; see equation (4c). For an impulse load, the qualitative dependence of the peak stress amplification/mitigation on granules' stiffness, mass density, and radius has been recognized in [4,21].

### 4.2.2. Error estimate when $L^P_{Applied} < A F_0^{5/6}$ and $t^P = nt_0$

In this case the first granule is always ejected, imparting additional momentum to the rest of the granules, which the calculation of $F_{SW_{max}}$ based on equation (5f) as outlined above, does not include. We now seek to estimate this momentum. Since the duration, $nt_0$, of the input force in equation (2) is measured in terms of its rise time, $t_0$, we first calculate a *virtual rise time*, say $\hat{t}_0$, such that, when the same input force-profile is measured in terms of this unit of time, $\hat{t}_0$, i.e., when the input force is $F(t; \hat{t}_0)$, it would generate a SW with $F_{SW_{max}} \simeq F_0$, or $L^P_{Applied} \simeq A F_0^{5/6}$. Hence $\hat{t}_0$ is given by,

$$\int_0^{\hat{t}_0} F_1(t)dt + \int_{\hat{t}_0}^{n\hat{t}_0} F_2(t)dt \simeq AF_0^{-1/6}, \tag{5g}$$

which can be calculated, for example, by iteration. The time-difference $\Delta t_0 = \hat{t}_0 - t_0$, is a measure of the error in the estimate that does not include the additional momentum due to an ejected granule. When $\Delta t_0 \simeq 0$, then the first granule would not be ejected, and if it should be then its linear momentum would be negligibly small. We estimate the linear momentum that would be generated if such ejection should occur, i.e. if $\Delta t_0 > 0$. The first granule begins to acquire a velocity anti-parallel with the applied force, when its contact area with the next granule begins to recede as the applied force decays. We use the terminal portion of the force $F(t;\hat{t}_0)$ (the shaded area in Figure 1), to estimate this additional momentum, $\Delta L \simeq 2 \int_{n\hat{t}_0-\Delta t_0}^{n\hat{t}_0} F(t;\hat{t}_0)dt$. Then, the total input momentum to form the SW would be $L_{Applied}^P + \Delta L \simeq L_{SW}$, from which we obtain,

$$F_{SW_{max}} \simeq \left( \frac{L_{Applied}^P}{A} + 2\frac{F_0}{A} \left\{ \int_{n\hat{t}_0-\Delta t_0}^{n\hat{t}_0} F_2(t;\hat{t}_0)dt \right\} \right)^{6/5}. \tag{5h}$$

Table 2 compares values of $F_{SW_{max}}$ given analytically by (5h) with the corresponding values obtain by numerical solution of equations (1). As is seen errors are quite small.

*4.2.3.    Estimation of $F_{SW_{max}}$ when $L_{Applied}^P \geq AF_0^{5/6}$ and $t^P = nt_0$*

In this case, the first granule may be ejected and either a single or double localized SWs can be generated; however the secondary SW would carry negligibly small energy. Therefore, it is reasonable to assume that all of the available input linear momentum is used to generate a single dominant SW wave, carrying almost all of the input momentum, and set $L_{Applied}^P \simeq L_{SW}$ which produces equation (5f). In this case, the calculation of $F_{SW_{max}}$ based on equation (5f) does not introduce significant errors when $L_{Applied}^P \geq AF_0^{5/6}$. This is due to the fact that even though the first granule (and possibly other granules, close to the application point of the input shock) may be ejected, the energy that they carry away would be insignificantly small.

### 4.2.4. Estimation of $F_{SW_{max}}$ when $t^P < nt_0$

In this case, multiple localized SWs are expected to form. First we estimate the percentage of the input momentum used to generate the 1$^{st}$ (primary) SW. This would serve to show whether or not all the input momentum is required to generate this SW.

Let the input momentum from time $t = 0$ to $t = t^P \left(= \alpha A F_{SW_{max}}^{-1/6}\right)$ be used to form a primary SW and set

$$\int_0^{\left(\alpha A F_{SW_{max}}^{-1/6}\right)} F(t;t_0) dt = \int_0^{\infty} F_{SW,1}(t) dt,$$

where $F_{SW,1}(t)$ represents the time variation of the primary SW. The linear momentum of the primary SW of peak amplitude $\max[F_{SW,1}(t)] = F_{SW_{max}}$ is $L_{SW,1} \simeq A F_{SW_{max}}^{5/6}$ and the linear momentum $L_{Applied}^P$ used to form the SW is given by $L_{Applied}^P = \int_0^{\left(\alpha A F_{SW_{max}}^{-1/6}\right)} F(t;t_0) dt \simeq L_{SW,1}$, which produces equation (5f).

### 4.2.5. Estimation of SWT amplitudes $F_{SW_{max},j}$ when $t^P < nt_0$

We now seek to calculate the number of SWs and their respective amplitudes. Let the input linear momentum from $t = 0$ to $t = \left(\alpha A F_{SW_{max},1}^{-1/6}\right)$ be used to form the primary SW, i.e., $\int_0^{\left(\alpha A F_{SW_{max},1}^{-1/6}\right)} F(t;t_0) dt = \int_0^{\infty} F_{SW,1}(t) dt$. Similarly, let the input momentum from time $t = \left(\alpha A F_{SW_{max},1}^{-1/6}\right)$ to $t = \left(\alpha A F_{SW_{max},2}^{-1/6}\right)$ be used to form the secondary SW, and so on. Then, in general, for the $j^{th}$ secondary SW, we have,

$$F_{SW_{max},j} \simeq \left[\frac{1}{A} \int_0^{\left(\alpha A F_{SW_{max},j}^{-1/6}\right)} F(t) dt - \sum_{k=2}^{j} F_{SW_{max},k-1}^{5/6}\right]^{6/5}; \qquad j = 1, 2, \cdots, \qquad (5i)$$

and hence,

$$\left(\alpha A F_{SW_{max},j}^{-1/6}\right) = \alpha A F_{SW_{max},1}^{-1/6} + \alpha_{New} A \sum_{k=2}^{j} F_{SW_{max},k}^{-1/6}; \quad j=1,2,\cdots; \text{ and}$$

$$\alpha_{New} = \frac{8-\sqrt{2}}{4} - \left|\frac{1}{2} - \frac{1}{(n-1)}\int_{1}^{n} F_2(t_0\tau)d\tau\right|^{3/2}.$$

Equation (5i) can be solved by iteration to obtain the peak values $F_{SW_{max},j}$ of the $j^{th}$ SW.

## 5. Discussion and Conclusion

When the input force to a chain of identical elastic spheres, has finite rise and decay durations, a rich set of interesting dynamic issues emerges. We have considered these issues for a general realistic class of shock-induced input force, and provided explicit expressions and procedures to successfully address them without having to numerically solve the required system of equations. Only the knowledge of the input force and the properties of the considered granules are required in our approach. For this, we have identified two fundamental parameters: one material-dependent parameter, $A$, and the other time parameter, $\left(\alpha A F_{SW_{max}}^{-1/6}\right)$, in addition to the input momentum parameter $L_{Applied}^P$. These parameters can be calculated *a priori* based on the properties of the granules and the time-profile of the input force. Then based on the relations between $L_{Applied}^P$ and $AF_0^{5/6}$, we have shown that:

i. If $L_{Applied}^P < AF_0^{5/6}$ and $t^P = nt_0$, then the first granule would be ejected and a single localized SW would be generated. As a result, the input energy would decrease and the input linear momentum would increase.

ii. If $L_{Applied}^P \geq AF_0^{5/6}$, then the first granule may be ejected, and at least one secondary SW would be generated.

iii. If $L_{Applied}^{P} \gg A F_0^{5/6}$, then the first granule would not be ejected and a train of SWs would be generated.
iv. When the system generates SWT, the number of SWs and their peak amplitudes can be calculated, as outlined in subsection 4.2.5.

In another word, we have shown for the first time that the entire physics of the SWs in our challenging problem is embedded in the values of only three fundamental quantities. In addition, and only from the structure of the parameter *A*, we have shown, *quantitatively,* how the peak value of SW, $F_{SW_{max}}$, increases with decreasing granules' radius, $R$, and density, $\rho$, and increasing Young's modulus, $E$. These observations have been made before by other researchers [4,21] qualitatively for an impulsive load only. Our equation (5f) provides new direct quantitative relation between the peak value of SW and the contacting granules' properties for finite duration shock loads.

While our focus has been on chains of identical elastic granules with one-directional Hertzian interaction, we hope that our methodology will pave a way to address similar challenges for two- and three-dimensional granular crystals subjected to general finite duration shocks.

**Acknowledgement**

This research has been conducted at the Center of Excellence for Advanced Materials (CEAM) at the University of California San Diego, under DARPA grant RDECOM W91CRB-10-1-0006 to the University of California, San Diego.

## References

[1] Deymier, P. A., 2013, Acoustic Metamaterials and Phononic Crystals, Springer Science & Business Media.

[2] Daraio, C., Nesterenko, V. F., Herbold, E. B., and Jin, S., 2006, "Tunability of solitary wave properties in one-dimensional strongly nonlinear phononic crystals," Phys. Rev. E, **73**(2), p. 026610.

[3] Nesterenko, V. F., 1983, "Propagation of nonlinear compression pulses in granular media," J. Appl. Mech. Tech. Phys., **24**(5), pp. 733–743.

[4] Nesterenko, V. F., 2001, Dynamics of Heterogeneous Materials, Springer.

[5] Leonard, A., and Daraio, C., 2012, "Stress Wave Anisotropy in Centered Square Highly Nonlinear Granular Systems," Phys. Rev. Lett., **108**(21), p. 214301.

[6] Hasan, M. A., Cho, S., Remick, K., Vakakis, A. F., McFarland, D. M., and Kriven, W. M., 2013, "Primary pulse transmission in coupled steel granular chains embedded in PDMS matrix: Experiment and modeling," Int. J. Solids Struct., **50**(20–21), pp. 3207–3224.

[7] Hasan, M. A., Cho, S., Remick, K., Vakakis, A. F., McFarland, D. M., and Kriven, W. M., 2014, "Experimental study of nonlinear acoustic bands and propagating breathers in ordered granular media embedded in matrix," Granul. Matter, **17**(1), pp. 49–72.

[8] Machado, L. P., Rosas, A., and Lindenberg, K., 2014, "A quasi-unidimensional granular chain to attenuate impact," Eur. Phys. J. E, **37**(11), pp. 1–7.

[9] Manjunath, M., Awasthi, A. P., and Geubelle, P. H., 2014, "Plane wave propagation in 2D and 3D monodisperse periodic granular media," Granul. Matter, **16**(1), pp. 141–150.

[10] Leonard, A., Ponson, L., and Daraio, C., 2014, "Exponential stress mitigation in structured granular composites," Extreme Mech. Lett., **1**, pp. 23–28.

[11] Hertz, H. R., 1988, "On the contact of elastic solids," J. Reine Angew. Math., **92**, pp. 156–171.

[12] Hong, J., 2005, "Universal Power-Law Decay of the Impulse Energy in Granular Protectors," Phys. Rev. Lett., **94**(10), p. 108001.

[13] Doney, R., and Sen, S., 2006, "Decorated, Tapered, and Highly Nonlinear Granular Chain," Phys. Rev. Lett., **97**(15), p. 155502.

[14] Daraio, C., Nesterenko, V. F., Herbold, E. B., and Jin, S., 2006, "Energy Trapping and Shock Disintegration in a Composite Granular Medium," Phys. Rev. Lett., **96**(5), p. 058002.

[15] Melo, F., Job, S., Santibanez, F., and Tapia, F., 2006, "Experimental evidence of shock mitigation in a Hertzian tapered chain," Phys. Rev. E, **73**, p. 041305.

[16] Harbola, U., Rosas, A., Esposito, M., and Lindenberg, K., 2009, "Pulse propagation in tapered granular chains: An analytic study," Phys. Rev. E, **80**, p. 031303.

[17] Machado, L. P., Rosas, A., and Lindenberg, K., 2013, "Momentum and energy propagation in tapered granular chains," Granul. Matter, **15**(6), pp. 735–746.

[18] Alves, M. K., and Alves, B. K., 1997, "No-tension elastic materials," Eur. J. Mech. Solids, **16**(1), pp. 103–116.

[19] Job, S., Melo, F., Sokolow, A., and Sen, S., 2007, "Solitary wave trains in granular chains: experiments, theory and simulations," Granul. Matter, 10(1), pp. 13–20.

[20] Starosvetsky, Y., and Vakakis, A. F., 2010, "Traveling waves and localized modes in one-dimensional homogeneous granular chains with no precompression," Phys. Rev. E, **82**(2), p. 026603.

[21] Nesterenko, V. F., 2002, "Shock (Blast) Mitigation by 'Soft' Condensed Matter in 'Granular Material-Based Technologies,'" MRS Symp. Proc., vol. 759 (MRS, Pittsburgh, PA, 2003), pp. MM4.3.1- 4.3.12 (arXiv:cond-mat/0303332v2).

**Figures**

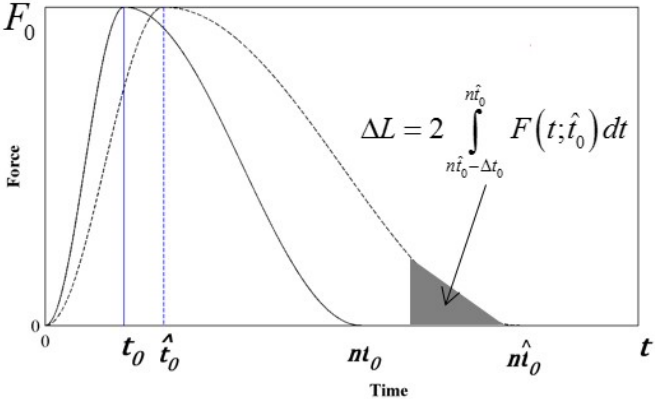

Figure 1. Schematic diagram showing the correction plot to estimate $F_{SW_{max}}$.

**Tables**

Table 1. Qualitative comparison between $F_{SW_{max}}$ and $F_0$ for *Power-law shock* input with $F_1(t)=(t/t_0)^{\eta_1}$ and $F_2(t)=\left[(nt_0-t)/\{(n-1)t_0\}\right]^{\eta_2}$. System parameters: $F_0=100N$, $t_0=15\mu s$ $E=193GPa$, $\rho=7958kg/m^3$, $D=9.5mm$.

| $(n,\eta_1,\eta_2)$ | Relation between $F_{SW_{max}}$ and $F_0$ | |
|---|---|---|
| | Exact (Numerical) | Approximate (Analytic) |
| (10,1,5) | $F_{SW_{max}} \simeq F_0$ | $F_{SW_{max}} \simeq F_0$ |
| (6,0.5,6) | $F_{SW_{max}} < F_0$ | $F_{SW_{max}} < F_0$ |
| (4,0.25,1) | $F_{SW_{max}} > F_0$ | $F_{SW_{max}} > F_0$ |

Table 2. Analytical estimation of peak value of the resulting primary SW $\left(F_{SW_{max}}\right)$ for *Power-law shock* input.

| $(n,\eta_1,\eta_2)$ | $(\hat{t}_0,t_0)$ in $\mu s$ | % of error in $F_{SW_{max}}$ using equation (5h) |
|---|---|---|
| (3,0.5,1) | (14.4, 7.4) | 3.0 |
| (2,1,1) | (24, 18) | 2.6 |
| (4,1e-5,1) | (9.6, 3.6) | 0.5 |
| (6,0.9,1) | (7.9, 1.9) | 2.6 |
| (5,0.25,1) | (8.6, 2.6) | 2.9 |